# Unconventional magnetic phase diagram of cuprate superconductor $La_{2-x}Sr_xCuO_4$ at quantum critical point x = 1/9

X. L. Dong[1], P. H. Hor[2*], F. Zhou[1], and Z. X. Zhao[1]

*1 National Lab for Superconductivity, Beijing National Laboratory for Condensed Matter Physics, Institute of Physics, Chinese Academy of Sciences, Beijing 100190, China*

*2 Texas Center for Superconductivity and Department of Physics, University of Houston, Houston, TX 77204-5002, USA*

**Abstract**

We propose a new magnetic phase diagram of $La_{2-x}Sr_xCuO_4$ around a quantum critical point x = 1/9 based on field-cooled magnetization measurements and critical fittings. A new phase boundary $T_{m2}(H)$ is discovered which buries deeply below the first order vortex melting line in the vortex solid phase. The coupling between superconductivity and antiferromagnetism is found to be attractive below $T_{m2}(H)$ while repulsive above. The attractive coupling between superconducting order and static antiferromagnetic order provides compelling experimental evidence that the antiferromagnetism microscopically coexists and collaborates with the high temperature superconductivity in cuprates.

---

[*]Corresponding author: phor@uh.edu

# 1. Introduction

Understanding how superconductivity (SC) emerges from the doped antiferromagnetic $CuO_2$ planes is the key to unlock the mystery of the mechanism for the high superconducting transition temperature ($T_c$) in cuprate superconductors. While doping holes into antiferromagnetic $CuO_2$ planes destroys magnetic order far before inducing superconductivity, at some specific doping concentrations and sample conditions static long range antiferromagnetic order can appear concomitantly with SC [1-4]. In these situations, upon applying a magnetic field (H) perpendicular to the $CuO_2$ planes, an H-induced enhancement of antiferromagnetism (AF) is ubiquitously observed.

The H-induced enhancement of AF was attributed to the proximity to a quantum critical point (QCP) where superconductivity and spin density wave (SDW) coexist microscopically, the SC+SDW phase. To the leading order the correction to the lowest magnetic energy mode in the SC+SDW phase is $\sim |v|[H/(2H_{c2})*\ln(\theta H_{c2}/H)]$ where v is the coupling constant between SC and SDW and $\theta = 3$ for the square/triangular vortex lattice [5]. The H-dependent enhancement of AF $\sim |v|(H/H_{c2})*\ln(H_{c2}/H)$ with v > 0 observed in refs [2, 6, 7] established that the amplitude of the SC order parameter (ψ) and that of the SDW order parameter (φ) are repulsively coupled, namely, SC and AF microscopically coexist and compete against each other in the (SC+SDW; v > 0) phase.

It was shown that the field-induced enhancement of antiferromagnetism arises from the coupling of the H to the currents circulating within the $CuO_2$ planes and magnetism contributes to superconductivity by neutron scattering experiment [8]. Instead of measuring H-dependence of magnetic response using neutron we focus on the H-dependence of the superconducting properties by measuring the field-cooled (FC) diamagnetic magnetization (M) of $La_{1.89}Sr_{0.11}CuO_4$ as a function of H and temperature (*T*). In this paper we report that, surprisingly, M exhibits the same H-dependence as that of the ordered moment of the coexisting antiferromagnetism. Furthermore $M^2$ at 5K increases with increasing field as $(H/(2H_{c2}))*ln(3H_{c2}/H)$ over four orders of magnitude in field from 4 Oersted (Oe) up to 1.3 Tesla. This result indicates a microscopically coexisting SC and SDW phase with an attractive coupling (*v* < 0), the (SC+SDW; *v* < 0) phase. For H > 1.3 Tesla, consistent with the neutron studies, we find that the coupling between SC and SDW is repulsive, a (SC+SDW; *v* > 0) phase. Therefore upon increasing H we cross a subtle *v*-transition from an attractive coupling (*v* < 0) between SC and AF to that of a repulsive (*v* > 0) one. This result is observed at one of the magic doping concentrations we identified before.

## 2. Experimental results and discussions

The magic doping concentrations are two dimensional hole densities $P_{MD} = m/n^2$ where m and n are positive integers with, for the two most prominent series, n = 3 or 4 and m ≤ n. A pristine electronic phase (PEP) is the electronic state of a cuprate superconductor doped at one of the magic doping concentrations. Many unusual physical properties and electronic phase separations of the PEPs have been observed in the single-layered $La_{2-x}Sr_xCuO_4$ (ref. [9]) and double-layered $YBa_2Cu_3O_{7-y}$ (ref. [10]), respectively. PEP has minimum complications due to electronic phase separations and will provide us with fundamentally clear and simple picture of the cuprate physics. We focus our study on one of the PEP with $P_{MD} = 1/3^2$ in $La_{2-x}Sr_xCuO_4$, namely, $La_{1.89}Sr_{0.11}CuO_4$. One of the most intriguing properties of PEPs relevant to the current study is the convergence of the FC diamagnetic susceptibility signal measured at 5 K ($^{5K}\chi_{FC}$) to that of the nearby magic doping concentration. Take $La_{2-x}Sr_xCuO_4$ for instance, upon increasing H, the $^{5K}\chi_{FC}$ of the x = 0.09 and x = 0.125 crystals, which are on either side of and close to $P_{MD} = 1/9$, started to "approaching" that of x = 1/9 when H > 0.1 Tesla. We attributed this behavior to the magnetic tuning with subsequent convergence of the electronic states of the x = 0.09 and x = 0.125 samples to the true energetically favored intrinsic vortex phase of the PEP at x = 1/9 (ref. [9]). Therefore $^{5K}\chi_{FC}$ is the intrinsic property of a PEP and so is the FC magnetization measured at each applied field.

In figure 1 we report the magnetic susceptibility (M/H) as a function of $T$ under various H. There are three characteristic temperatures that we can visually identify in each $4\pi M(T)/H$ curve: (1) the onset of superconducting transition temperature $T_{C,O}(H)$ defined as the temperature when magnetization versus temperature curve starts to decrease; (2) a first order melting transition $T_1$ is clearly marked by a magnetization jump ($\triangle M$) and (3) the irreversibility temperature $T_{irr}$ can be easily identified by the separation between the FC and zero-filed-cooled (ZFC) magnetic susceptibility versus temperature curves.

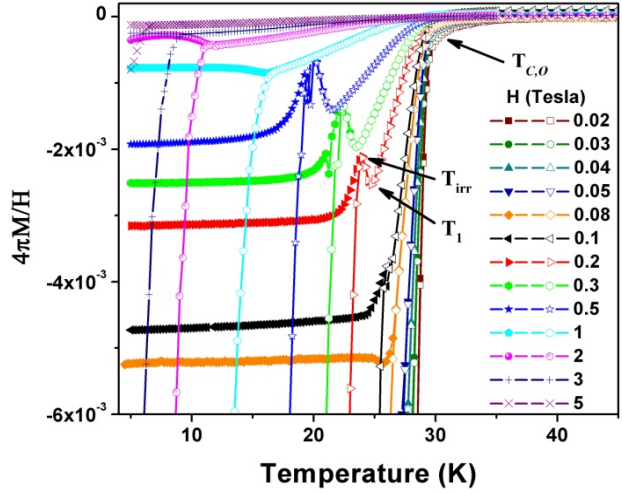

**Figure 1:** Magnetic susceptibility as a function of temperature of a $La_{1.89}Sr_{0.11}CuO_4$ crystal under ZFC (open) and FC (solid) conditions. The H is applied perpendicular to the ab-plane. Various characteristic temperatures can be visually identified (see text for details).

In figure 2 we plot the two characteristic temperatures $T_1$ and $T_{irr}$ identified in figure 1 together with the first order melting transition temperature $T_{FOT}$ and the irreversibility temperature $T_{irr}'$ determined by the ac-susceptibility and magnetization measurements reported in Ref. [11]. The irreversibility temperature $T_{irr}''$ determined by the muon-spin rotation and small-angle neutron scattering experiments [12] is also included in the figure 2. We can see that the two melting temperatures $T_1$ and $T_{FOT}$ and the three irreversibility temperatures $T_{irr}$, $T_{irr}'$ and $T_{irr}''$ are all very consistent with each other in spite of the fact that these characteristic temperatures are determined by three different groups using four different experimental probes on separately prepared crystals. What is common among them is the close proximity to the magic doping concentration x = 1/9.

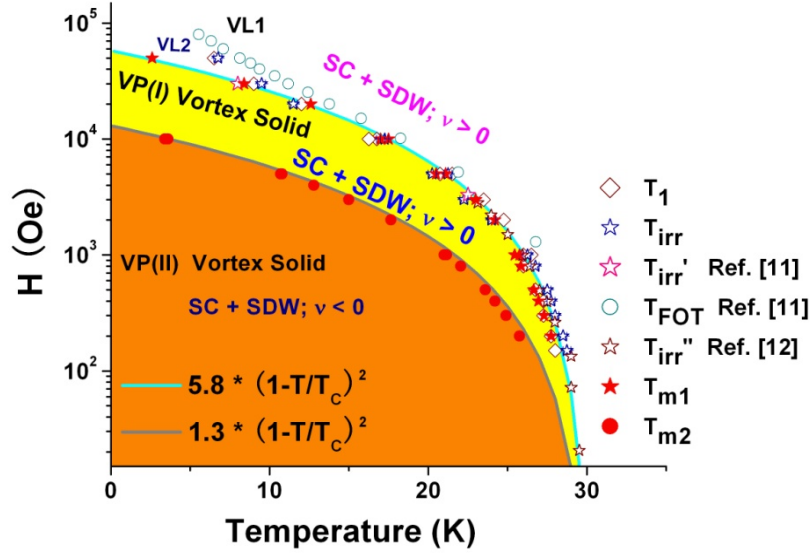

**Figure 2:** Magnetic phase diagram of $La_{1-x}Sr_xCuO_4$ crystals around x = 1/9. Various characteristic temperatures identified in figure1 ($T_1$ & $T_{irr}$), reported in the literatures ($T_{irr}$', $T_{irr}$'' & $T_{FOT}$) and extracted by the critical fittings ($T_{m1}$ & $T_{m2}$) are plotted together for comparison. $T_{m1}$(H) & $T_{m2}$(H) are the boundaries that separate VL from VP(I) and VP(I) from VP(II) phases, respectively. At low temperature (< 15K) and under high field ( > 2Tesla) there seems to have two different VL states, VL1 & VL2, where VL2 is the completely melted vortex liquid state and the nature of VL1 is currently unknown.

In figure 3a we show that 2D critical fluctuations of the scaling variable x ~ [$T-T_c$(H)]/($T$H)$^{1/2}$ (refs [13, 14]) fits extremely well near $T_{C,O}$ up to 5 Tesla. The fitted region grows with increasing H and the entire reversible magnetization can be scaled together for H > 1 Tesla. Noted that there is no fitting parameter and the fitted transition temperature $T_{m1}$ (closed (red) stars in figure 2) closely corresponds to the $T_{FOT}$ and irreversibility temperatures up to 2 Tesla. For H > 2 Tesla there seems to have two different vortex liquid (VL) phases, VL1 and VL2, defined by different experimental probes. The origin of the VL1 and VL2 is not clear to us. We conclude that $T_{m1}$ is the well-known melting transition of vortex solid into VL states.

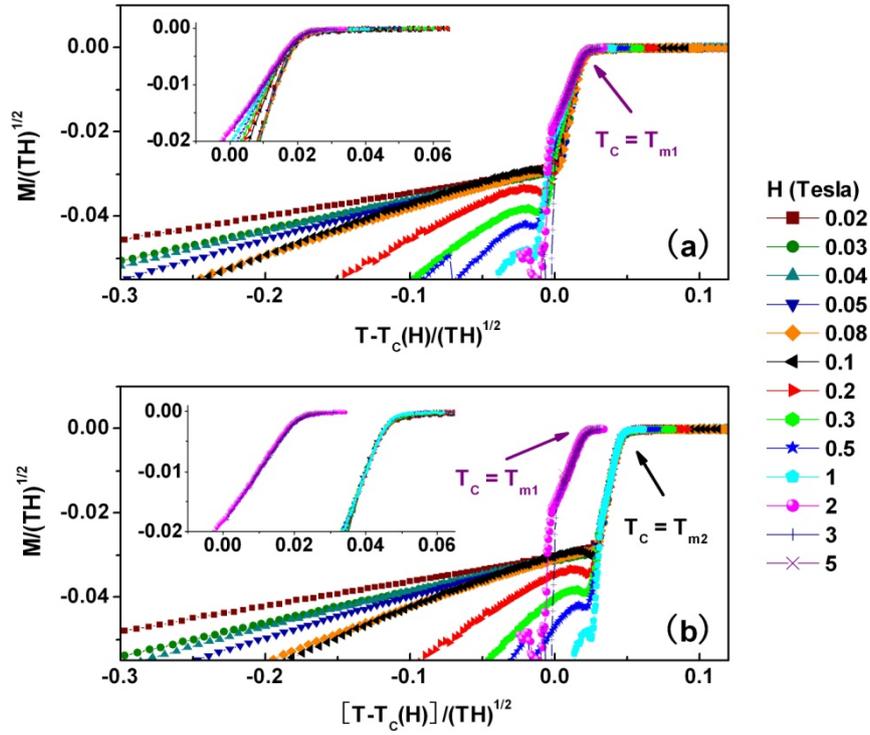

**Figure 3:** 2D critical fitting of magnetization as a function of temperature of the $La_{1.89}Sr_{0.11}CuO_4$ crystal. (a) Fittings over two orders of magnitude in field (0.02 Tesla ≤ H ≤ 5 Tesla) yield $T_{m1}$. $T_{m1}$ is extracted from a region from onset to ~30% below the onset temperature. The inset is an expanded view of the fitting. (b) $T_{m2}$ is extracted by 2D critical fitting (right) to the entire magnetization vs. temperature curve up to 1 Tesla. The fitting on the left includes entire magnetization vs.

temperature curve for H > 1 Tesla which yields the corresponding part of the transition line $T_{m1}$ above one Tesla. The inset is an expanded view of the fitting.

On the other hand we can also scale the entire reversible magnetization together as shown to the right in figure 3b. It works up to 1 Tesla and the fitted transition temperature $T_{m2}$ (closed (red) circles in figure 2) does not correspond to any previously identified transition in figure 2. It indicates that, for H < 1 Tesla, the critical fluctuation is dominated by the transition to the phase below $T_{m2}$ and, for H > 1 Tesla, the critical fluctuation is dominated by the transition to the phase below $T_{m1}$. Therefore the reversible magnetization below $T_{C,O}$ is not due to a single superconducting transition, it composes of critical fluctuations of two phase transitions toward the phases below $T_{m1}$ and $T_{m2}$.

We need to emphasize that our critical fittings to extract $T_{m1}$ and $T_{m2}$ worked only for x = 1/9. It is a unique property of the PEP. We cannot obtain $T_{m1}$ and $T_{m2}$ from x = 0.09 or 0.125 crystals since neither of their reversible magnetization data show scaling behavior for H < 1 Tesla. However 2D critical fitting worked for both x = 0.09 and 0.125 crystals for H > 1 Tesla which yields the same $T_{m1}$ as that of x = 1/9. This result is consistent with and confirms our previous conclusion that both crystals entered the intrinsic vortex phase of x =1/9 at 5K for 1 Tesla < H < 5 Tesla [9].

Both phase boundaries $T_{m1}$(H) and $T_{m2}$(H) can be well described by $H_m \sim H_0(1-T/T_c)^2$ （cyan and gray lines in figure 2）where $H_0$ = 5.8 Tesla and 1.3 Tesla, respectively. This is the well-known H-dependence of a vortex melting line for clean superconductors [15]. Since both $T_{m1}$(H) and $T_{m2}$(H) are extracted from critical fittings to the reversible magnetization and behave as a melting type transition we conclude that $T_{m1}$(H) and $T_{m2}$(H) are true equilibrium phase boundaries and the vortex phase I (VP(I)) between phase boundaries $T_{m1}$(H) and $T_{m2}$(H) and vortex phase II (VP(II)) below $T_{m2}$($H_{m2}$) line are two true intrinsic thermodynamic phases. From the plot in figure 2 and fitting range used we can clearly see that both $T_{m1}$ and $T_{m2}$ transitions are suppressed by the 2D fluctuations to a substantially lower temperature. At 1 Tesla, for instance, the fitted fluctuation region starts from 30K and ends at 4K and 15K above $T_{m1}$ and $T_{m2}$, respectively.  While $T_{m1}$(H) can be easily understood as the first order melting transition from a vortex solid into a vortex liquid, the VP(I), VP(II) and the $T_{m2}$(H) are quite unusual and intriguing. Both VP(I) and VP(II) phases are located within the vortex solid phase and there is no other thermodynamic vortex phases known in this region from previous studies, both theoretically and experimentally. Therefore, $T_{m2}$(H) line is a subtle and yet-to-be-identified phase transition in the phase diagram.

We will address the VP(I), VP(II) phases and the $T_{m2}$(H) boundary based on the theoretical treatment of the near quantum critical SC+SDW phase [5].  We argue that all the samples we have discussed here are very close to a QCP and is intimately related the SC+SDW phase. Furthermore we postulate that our sample at magic doping concentration x= 1/9 is sitting almost right on the QCP and the two dimensional hole densities *P* is the tuning parameter for the quantum phase transition. Other factors such as dopant distribution, lattice imperfections and impurities will also displace the electronic state away from the QCP. However we consider all the above effects are very small in our high quality $La_{1.89}Sr_{0.11}CuO_4$ crystal. Therefore how far a sample is away from the QCP, namely, $r = P-1/9$ is a measure of the proximity to the QCP for a sample with doping concentration *P*. In order to compare with the neutron experiments and since we find that $^{5K}\chi_{FC}$ is an intrinsic thermodynamic quantity of the x =1/9 crystal [9] we now formally define an *effective diamagnetic moment* $M_{eff}$= M/number of Cu atoms per unit volume for each Cu atom. M is the FC diamagnetic moment per unit volume. Essentially we treat SC+SDW as a thermodynamic phase with a diamagnetic moment $M_{eff}$ per copper atom.

Using $H_{c2}$ and ν as the fitting parameters we find that at 5K, as shown in figure 4a, $M_{eff}^2 = |\nu|(H/(2H_{c2}))*\ln(3H_{c2}/H)$ for 4 Oe < H < 1.3 Tesla with $|\nu|$ = 0.000085 and $H_{c2}$ = 1.3 Tesla. Note that the fitted $H_{c2}$ is only 1.3 Tesla which is much smaller than the upper critical field ~ 60 Tesla [16]. Physically, it should be interpreted as the critical field of the superconducting phase below $T_{m2}$ phase boundary. Indeed, $H_{c2}$ = 1.3 Tesla coincides with the melting-like critical field $H_0$ obtained by the 2D fluctuation scaling fitting as shown in figure 2. In the inset of figure4a we plot the $M_{eff}^2$ versus lnH so we can clearly compare and see the quality of the fittings both in the linear and

logarithmic scales. This highly non-trivial H-dependence of $M_{eff}^2$ over almost four orders of magnitude in H is one of the key results in this report and, again, it only works for the x = 1/9 crystal. It cannot be observed in x = 0.09 and x = 0.125 crystals. Note that the $M_{eff}^2 \sim |\nu|(H/(2H_{c2}))*\ln(3H_{c2}/H)$ dependence starts at such a low field ~ 4 Oe is a further confirmation that our crystal is "very" close to the QCP.

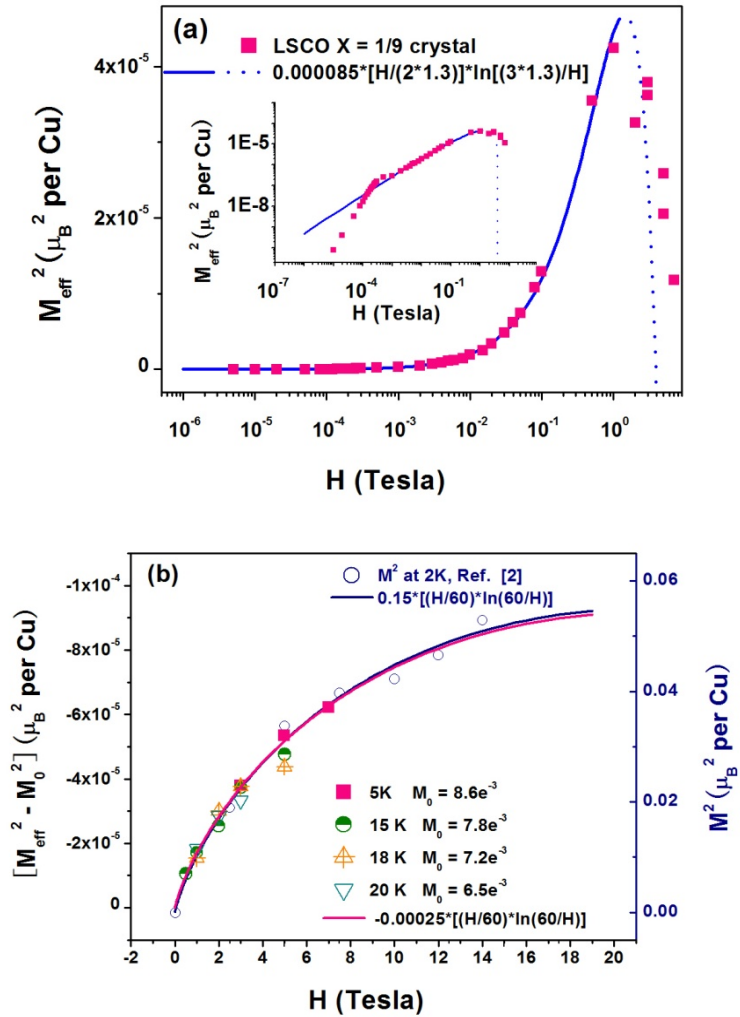

**Figure 4:** $M_{eff}^2$ vs. H of $La_{1.89}Sr_{0.11}CuO_4$ crystal. (a) For $\nu < 0$, the $M_{eff}^2$ vs. H curved at 5K is fitted to the field dependence of $M_{eff}^2 \sim |\nu|(H/(2H_{c2}))*\ln(3H_{c2}/H)$ of the microscopically coexisted superconductivity and spin density wave state. The solid

and dotted curves correspond to the magnetic field regime below and above $T_{m1}$ transition line, respectively. The inset is the ln-ln plot of the data to show an alternate view of the fit. (b) For $\nu > 0$, we compare the H-dependence of diamagnetic moment to that observed by neutron scattering experiment. Since the diamagnetic moment decreases with increasing H for H> $T_{m2}$ at a fixed $T$ we fit $M_{eff}$ as $(M_{eff}^2-M_0^2) = -\nu(H/H_{c2})*\ln(H_{c2}/H)$ at 5K, 15K, 18K and 20K where $M_0$ is a fitting constant for each temperature (left scale). The pink solid curve is our fitting of $(M_{eff}^2 - M_0^2) = -\nu(H/(H_{c2}))*\ln(H_{c2}/H)$ with $H_{c2}$ = 60 Tesla. The fitted $\nu$ = 0.00025 Bohr magneton (left scale). The blue open circles are the square of the field-induced enhancement of magnetic moment ($M^2$) per $Cu^{++}$ ion in Bohr magneton (right scale) measured at 2K by neutron scattering experiment. The blue solid curve is our fitting of $M^2 \sim \nu(H/(H_{c2}))*\ln(H_{c2}/H)$ with $H_{c2}$ = 60 Tesla. The fitted $\nu$ = 0.15 Bohr magneton (right scale).

In figure 4a the slope of $M_{eff}^2$ vs. H curve changes from positive to negative around H = 1.3 Tesla. Since $M_{eff}$ is diamagnetic, so $dM_{eff}/dH > 0$ corresponds to a negative coupling constant $\nu$ and $dM_{eff}/dH < 0$ implies $\nu > 0$. We can see that $dM_{eff}/dH > 0$ and $dM_{eff}/dH < 0$ exist in a field region that is above and below $T_{m2}(H)$ phase boundary line, respectively. Therefore we conclude that the coupling constant $\nu$ should be negative for the phase below the $T_{m2}(H)$ line and positive above. Therefore the $T_{m2}(H)$ phase boundary is related to a subtle change from a microscopically coexisting and co-operating superconducting and spin density wave state (SC+SDW; $\nu < 0$) to a microscopically coexisting and competing superconducting and spin density wave state (SC+SDW; $v > 0$).

In figure 4b we show that the phase with competing SC and SDW, the (SC+SDW; $v> 0$) phase above the $T_{m2}$(H) line, is indeed quantitatively consistent with that obtained by the neutron scattering experiment [2]. We plot $M_{eff}^2$ vs. H at various constant $T$. The pink line is the fitting to $(M_{eff}^2 - M_0^2) = -v(H/H_{c2})*\ln(H_{c2}/H)$ at a fixed temperature using $H_{c2}$ = 60 Tesla [16]. The $v = 0.00025$ is the repulsive coupling constant and $M_0^2$ is a temperature dependent constant offset. The navy circles are the square of the field-induced magnetic moment per $Cu^{++}$ ion measured at 2K by neutron scattering experiment [2]. The solid navy curve is our fitting to $M^2 = v(H/H_{c2})*\ln(H_{c2}/H)$ to the neutron data extracted from Ref. 2 with $v = 0.15$ using $H_{c2}$ = 60Tesla. We can clearly see that all sets of data overlap together, indicating that the field-induced *decrease* of diamagnetic moment is the same as that of the field-induced *increase* of the magnetic moment observed by neutron scattering experiments. This confirms that (SC+SDW; $v> 0$) is indeed a coexisting SC and AF phase with repulsive interaction. Note that $v = 0.00025$ is much smaller than $v = 0.15$ from neutron fitting. This indicates that diamagnetic moment cannot be the only source for H-induced enhancement of magnetism, there are yet to be identified other contributions.

We find that for a good fitting it is necessary to use $\theta = 3$ in the $\ln(\theta H_{c2}/H)$ term in $M_{eff}^2 \sim |v|(H/(2H_{c2}))*\ln(3H_{c2}/H)$ to extract the (SC+SDW; $v< 0$) phase. Therefore we believe well-ordered square/triangular lattices exist in the (SC+SDW; $v<0$) phase. However if $\theta=3$ was used to extract (SC+SDW; $v>0$) phase it will yield a worse fitting with un-physical $H_{c2} \sim 20$ Tesla. Therefore, we conclude that there is a highly disordered vortex solid in the (SC+SDW; $v>0$) phase. Therefore the coupling between SC and SDW is attractive in the conventional Abrikosov flux lattice. It is repulsive in the disordered vortex solid that extends into vortex liquid phase.

The existence of the (SC+SDW; $v<0$) phase provides a direct experimental support to the magnetic contribution to high $T_c$ in cuprates since the diamagnetic signal we measured is due to circulating currents in the $CuO_2$ planes. It also naturally explains, when away from magic doping, perturbations such as magnetic field or high pressure will push the electronic system to a nearby energetically favored intrinsic $T_c$ at a magic doping concentration [9, 17]. In the (SC+SDW; $v>0$) phase the antiferromagnetism extends $\sim 6d$ along c-axis [8] and SC usually has a coherence length $< d$ where $d \sim 6.6$Å is the inter-planar distance between $CuO_2$ planes. Therefore, strictly speaking, neither superconductivity nor antiferromagntism exhibits a true 3D long range order in the (SC+SDW; $v>0$) phase; they coexists at the expense of a limited extension along c-axis direction. It is possible that they are also limited in size in the $CuO_2$ planes although it was shown that they coexisted with much larger size $\geq 400$Å which is the resolution limit of the neutron experiment [2].

To place our results into a broader prospect, in light of the very generic nature of the theory, other similar orders such as charge density wave or "staggered flux" phases may also attractively couple to superconductivity and produce phase diagrams similar to that presented in figure 2. While we discuss our results entirely based on the antiferromagnetic order that appears in the superconducting state we need to point out that the recent observation of an unusual magnetic order at pseudogap temperature [18] should also play a role here. More details neutron experiments perform at low field (< 5 Tesla) are necessary to address this issue.

## 3. Conclusion

In summary, we present the unconventional magnetic phase diagram of $La_{1.89}Sr_{0.11}CuO_4$ at QCP which includes the new novel (SC+SDW; $v<0$) phase in the vortex solid state. By performing critical scaling analysis of reversible magnetization data measured on $La_{1.89}Sr_{0.11}CuO_4$, we extracted two phase boundaries $T_{m1}(H)$ and $T_{m2}(H)$ that correspond to the first order melting transition and a very subtle $v$-transition, respectively. In contrast to the common belief that superconductivity and antiferromagnetic orders are always competing against each other in cuprates, the superconductivity and antiferromagnetism are actually co-operating with each other in the (SC+SDW; $v<0$) phase below $T_{m2}(H)$. Our observation of this new low energy attractive interaction between SC and AF underlies the cuprate physics. It unifies and puts the superconductivity and antiferromagnetism on equal footing which, in turn, provides a fundamental new direction to construct the microscopic theory for high $T_c$ cuprates. For instance, while a uniformly mixed repulsive SC and AF orders is a natural consequence in the SO(5) theory [19] it remains to see if SO(5) theory can also generate a microscopically coexisting phase with co-operating SC and AF order. Further studies are necessary to pin down the microscopic origin of the attractive interaction between superconductivity and antiferromagnetism.

**Acknowledgement**

We thank Prof. F.C. Zhang for valuable discussions. We also thank the members in Z.X.Z's group and P.H.H's group for assistance with sample preparation and characterization. This work is supported by the NSF of China (project Nos. 10874211) and the National Basic Research Program of China (project Nos. 2007CB925002 & 2011CB921703) and the State of Texas through the Texas Center for Superconductivity at University of Houston.